# On the Possible Thermal *Tachyons*

Miroslaw Kozlowski[*]
Janina Marciak-Kozlowska

Institute of Electron Technology
Al. Lotników 32/46. 02-668 Warsaw, Poland

---
[*]Corresponding author, e-mail: kozlo@ite.waw.pl

Abstract
In this paper the existence of the thermal *tachyons* i.e. quanta of temperature field, with $v > c$ is described in the theoretical frame of hyperbolic thermal equation. The modified Lorentz transformation are developed. It is argued that thermal *tachyons* can exist in accordance with modified Lorentz transformation after change $c^2 \to -c^2$. The thermal *tachyons* fulfill the hyperbolic heat transport equation and in principle can be created by attosecond laser pulses.

# 1. Introduction

The square of the neutrino mass was measured in tritium beta decay experiments by fitting the shape of the beta spectrum near endpoint. In many experiments it has been found to be negative. According to the results of paper [1]

$$m^2(v_e) = -2.5\,\text{eV}^2.$$

Based on special relativity superluminal particles, i.e. particles with $m^2 < 0$ and $v > c$ were proposed and discussed in paper [1, 2]. In this paper we investigate the possibility of the existence of the superluminal particles from the point of view the hyperbolic heat transport equation. It will be shown that hyperbolic heat transport equation is invariant under transformation $c^2 \to -c^2$, i.e. for transformation $c \to ic$. The new Lorentz transformation for $-c^2$ and formula for kinetic energy will be developed. It will be shown that for $E_k/mc^2 > 1$ the speed of particles is decreasing for increased kinetic energy $E_k$.

# 2. The Equation

In our monograph the hyperbolic Heaviside transport equation for attosecond laser pulses was obtained [3]

$$\frac{1}{v^2}\frac{\partial^2 T}{\partial t^2} + \frac{1}{D_T}\frac{\partial T}{\partial t} = \nabla^2 T. \tag{1}$$

In this equation $T$ is the absolute temperature, $v$ denotes the speed of the thermal disturbance, and $D_T$ is diffusion coefficient for thermal phenomena. Equation (1) describes the damped thermal wave propagation. Recently, the observation of the thermal wave in GaAs films exposed to ultra-short laser pulses was presented [2]. In the subsequent we will consider the mathematical structure of the Eq. (1) which can be interested for both the experimentalist as well as the theorist involved in ultrahigh ultra-short laser pulses. In the monograph [3] it was shown that speed $v$, and diffusion coefficient $D_T$ in Eq. (1) can be written as

$$v = \alpha c,$$
$$D_T = \frac{\hbar}{m}. \tag{2}$$

In formula (2) $\alpha$ is the electromagnetic fine structure constant, $\hbar$ is the Planck constant, $m$ is the mass of the heat carrier, and $c$ is light velocity.

Let us consider the transformation

$$c \to ic \tag{3}$$

for the Eq. (1). First of all we note that

$$v'^2 \to \alpha^2 c^2 = v^2,$$
$$D_T' = D_T. \tag{4}$$

We conclude that the Eq. (1) is invariant under the transformation (3). One can say that Eq. (1) is valid for the universe for which $-c^2$ is the invariant constant.

## 3. The Lorentz Transformation for $-c^2$ Universe

In our Universe the Lorentz transformation has the form

$$x' = \frac{x - vt}{\sqrt{1 - \frac{v^2}{c^2}}},$$

$$t' = \frac{t - \frac{v}{c^2} x}{\sqrt{1 - \frac{v^2}{c^2}}}. \tag{5}$$

With the transformation (3) we obtain from formula (5)

$$x' = \frac{x - vt}{\sqrt{1 + \frac{v^2}{c^2}}},$$

$$t' = \frac{t + \frac{v}{c^2} x}{1 + \frac{v^2}{c^2}}. \tag{6}$$

For the space-time interval we obtain

$$x^2 + c^2 t^2 = (x')^2 + c^2 (t')^2$$

as in $c^2$ special relativity.

For $-c^2$ SR we have new formula for the velocities

$$v' = \frac{v - V}{1 + \frac{vV}{c^2}} \tag{7}$$

For speed $v = ic$, from formula (7) we obtain $v' = ic$, i.e. object with speed $ic$ has the same speed in all inertial reference frames. Now, we consider the formulae for total energy and momentum of the particle with mass $m$,

$$E = \frac{-mc^2}{\sqrt{1 + \frac{v^2}{c^2}}}, \qquad p = \frac{mv}{\sqrt{1 + \frac{v^2}{c^2}}}. \tag{8}$$

From formula (8) we obtain

$$\frac{E}{p} = -\frac{c^2}{v}. \tag{9}$$

For objects with speed $v = ic$ we obtain from formula (9)

$$\frac{E}{p} = ic. \tag{10}$$

Considering formula

$$E = \sqrt{-p^2c^2 + m^2c^2} \tag{11}$$

we obtain, for $m = 0$

$$E = ipc, \quad \frac{E}{p} = ic. \tag{12}$$

We conclude that objects with masses $m = 0$, have speed $v = ic$ in $-c^2$ universe. We calculate the kinetic energy, $E_k$

$$E_k = -mc^2(\gamma - 1), \quad \gamma = \frac{1}{\sqrt{1 + \frac{v^2}{c^2}}}. \tag{13}$$

From formula (13) we deduce the ratio $v^2/c^2$

$$\frac{v^2}{c^2} = \frac{1 - \left(1 - \frac{E_k}{mc^2}\right)^2}{\left(1 - \frac{E_k}{mc^2}\right)^2}. \tag{14}$$

It is quite interesting that in $-c^2$ universe $v^2/c^2$ is singular for $E_k = mc^2$, i.e. when kinetic energy of the objects equals its internal energy. For the $c^2$ universe the formula which describes $(v/c)^2$ reads

$$\frac{v^2}{c^2} = \frac{\left(1 + \frac{E_k}{mc^2}\right)^2 - 1}{\left(1 + \frac{E_k}{mc^2}\right)^2}. \tag{15}$$

## 4. Thermal *Tachyons*

In monograph [3] we show that the Eq. (1) describes the propagation of heatons, quanta of thermal field $T(x,t)$. For quantum heat transfer equation (1) we seek solution in the form

$$T(x,t) = e^{-\frac{t}{2\tau}} u(x,t). \tag{16}$$

After substitution of Eq. (16) into Eq. (1) one obtains

$$\frac{1}{v^2} \frac{\partial^2 u}{\partial t^2} - \frac{\partial^2 u}{\partial x^2} + qu(x,t) = 0, \tag{17}$$

where

$$q = -\left(\frac{mv}{2\hbar}\right)^2. \tag{18}$$

It is interesting to observe that if we introduce the imaginary mass $m^* = im$ then formula (18) can be written as

$$q = \left(\frac{m^* v}{2\hbar}\right)^2, \tag{19}$$

and Eq. (17) is the Klein-Gordon for *heatons* with imaginary mass $m^* = im$. According to the results of our monograph[3] we can call *heatons* with imaginary mass $m^*$ the *tachyons*, i.e. particles with $v > c$.

In Figs. 1, 2 we present the calculation of the ratio $\left(v/c\right)^2$ according to the formula (14). As can be seen only for $E_k/mc^2 \ll 1$, both formulae (14) and (15) give the same results. We can conclude that in the region of $E_k/mc^2 > 1$ we can, in principle, discover the strange behavior of the particle, with increasing $E_k$ its speed is decreasing.

One can conclude that the irradiation of the matter with attosecond laser pulse can produce the thermal *tachyons*, which propagate with speed $v > c$. It must be stressed that the fact that $v > c$ does not violate the special relativity as described in paragraph 1. Moreover, the *thermal tachyons* fulfils the same thermal transport equation (1) as the particles with $v < c$. The experimental method best suited for the observation of the thermal *heatons* is the TOF (Time of Flight) measurement of the velocity of the emitted particles.

On the experimental ground the existence of the *tachyons* is still an open question. However, the results of papers [1, 2] strongly suggest the existence of the particles with masses $m^* = im$, i.e. $\left(m^*\right)^2 = -m^2 < 0$.

## Conclusions

In this paper it is shown that the hyperbolic heat transport equation can be applied to the study of the *thermal tachyons*. It is argued that the particles with masses $m^* = im$ can propagate with velocity $v > c$ and do not break the new Lorentz transformation

# References

[1] T. Chang, *Nucl. Sci. Tech*. **13**, (2002) 129.

[2] X. Ai, B.Q. Li, *Journal of Elec. Mat*. **34**, (2005) 583.

[3] M. Kozlowski, J. Marciak-Kozlowska, *Thermal Processes Using Attosecond Laser Pulses*, Springer, 2006.

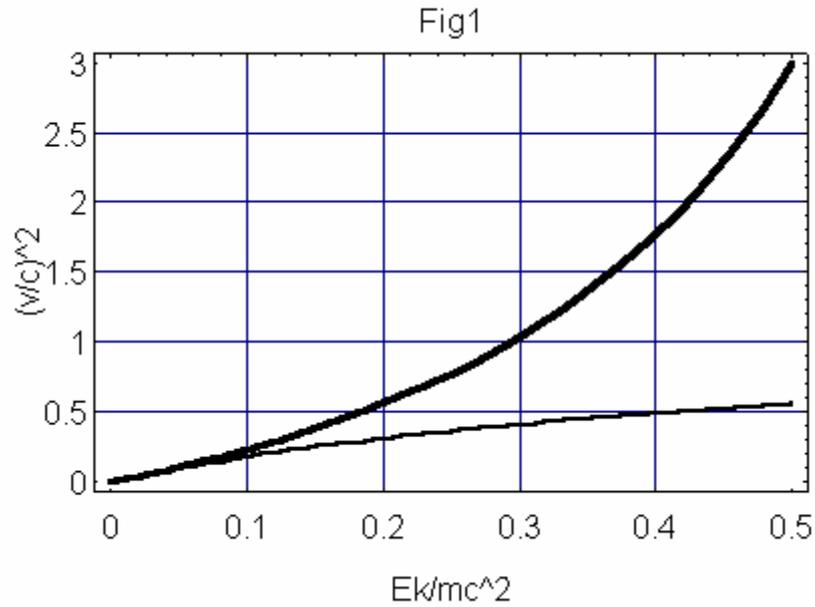

**Fig. 1.** The $\left(v/c\right)^2$ as the function of the ratio $E_k/mc^2$. Only for $E_k/mc^2 < 0.2$ both description with $c^2$, and $-c^2$ give the same results. Thick curve, formula (14), thin - formula(15)

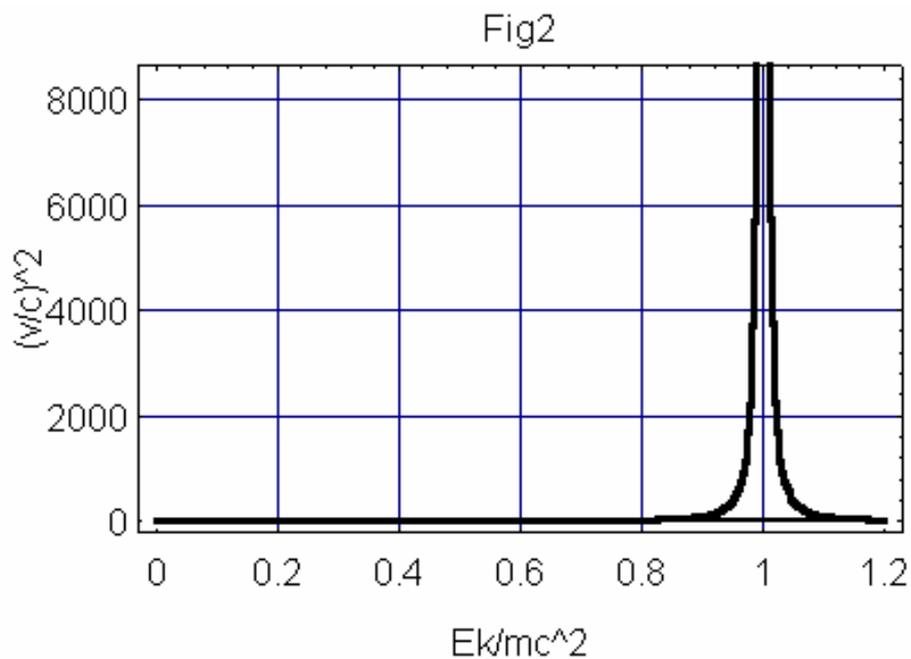

**Fig 2.** For $E_k/mc^2 \to 1$, $\left(v/c\right)^2 \to \infty$. For $E_k/mc^2 > 1$, the speed of the particles is decreasing with increased kinetic energy.